\documentstyle[12pt,aasms4]{article}

\newcommand{\teff}{$T_{\rm eff}$}
\newcommand{\kms}{\mbox{${\rm km~s}^{-1}$}}
\newcommand{\ha}{\mbox{H$\alpha$}}

\newcommand{\vrad}{$v_{\rm rad}$}
\renewcommand{\bv}{\mbox{$(B-V)$}}

\newcommand{\vsini}{\mbox{$v\,\!\sin\,\!i$}}

\newcommand{\lR}{\mbox{$\log R_{\rm HK}^\prime$}}
\newcommand{\FH}{$F_{{\rm H}\alpha}$}
\newcommand{\RH}{$R_{{\rm H}\alpha}$}

\newcommand{\ALi}{\mbox{$\log N({\rm Li})$}}

\newcommand{\etal}{et al.}
\begin{document}
\title{High-Resolution Spectroscopy of Some Very Active Southern Stars}
\author{David R. Soderblom\altaffilmark{1}, Jeremy R. King}
\affil{Space Telescope Science Institute\\
3700 San Martin Drive, Baltimore MD 21218} 
\authoremail{soderblom@stsci.edu, jking@stsci.edu}

\and

\author{Todd J. Henry}
\affil{Harvard-Smithsonian Center for Astrophysics,\\
60 Garden Street, Cambridge MA 02138}
\authoremail{thenry@cfa.harvard.edu}

\altaffiltext{1}{Visiting astronomer, Cerro Tololo Inter-American Observatory,
which is operated by the Association of Universities for Research in Astronomy,
Inc., under contract with the National Science Foundation.}

\begin{abstract}
We have obtained high-resolution echelle spectra of 18 solar-type stars that
an earlier survey showed to have very high levels of \ion{Ca}{2} H and K
emission.  Most of these stars belong to close binary systems, but 5 remain as
probable single stars or well-separated binaries that are younger than the
Pleiades on the basis of their lithium abundances and \ha\ emission.  Three of
these probable single stars also lie more than 1 magnitude above the main
sequence in a color-magnitude diagram, and appear to have ages of 10 to 15
Myr.  Two of them, HD 202917 and HD 222259, also appear to have a kinematical
association with the pre-main sequence multiple system HD 98800.

\end{abstract}

\section{Introduction}

Two years ago we published a survey of \ion{Ca}{2} H and K emission strengths
in more than 800 southern solar-type stars (\cite{h96} 1996; hereafter
Paper I), determined from low-resolution ($R\approx2,000$) spectra obtained at
CTIO.  The purpose of the survey was to provide at least rough estimates of
the ages of the individual stars, and to examine the distribution of emission
strengths in a large and unbiased sample.  The very existence of a simple or
single relationship between chromospheric emission (CE) and age is debatable,
but there is ample evidence that CE declines steadily with age (\cite{sk72}
1972; \cite{sdj} 1991).  In other words, we are confident of a general decline
of CE with age because of observations of stars in clusters, and also because
CE is so intimately tied to stellar rotation and we know that rotation
declines with age in solar-type stars.  But we also know that the CE levels
of individual stars vary due to rotational modulation of active regions,
long-term cycles, and other phenomena, and we also know that not all stars
reach the Zero-Age Main Sequence (ZAMS) with the same angular momentum, and
thus we are not so sure that there is a unique CE-age relation that applies
to all stars.

The survey of Paper I included stars from a G-dwarf sample defined using the
combination of two-dimensional spectral types (\cite{hk1} 1975;
\cite{hk2} 1978; \cite{hk3} 1982; \cite{hk4} 1988) and Str\"omgren
photometry (\cite{o88} 1988, \cite{o93} 1993), and it turned up two groups of
stars that we have studied further.  The first group we called ``Very Active''
because they exhibited CE levels well above any seen in the field stars of the
earlier survey of \cite{vp80} (1980).  The second group is called ``Very
Inactive'' and consists of stars that appear to have activity levels well below
the Sun's.  The Very Inactive stars will be the subject of a future paper. 

Here we concentrate on the Very Active stars, and we are led to examine them in
detail for several reasons.  First, stars with very high levels of activity are
that way because they rotate rapidly, and they rotate rapidly either because
they are very young (and have not yet lost much of their initial angular
momentum), or because they are in a close binary system (where the companion's
tidal forces lead to synchronous rotation).  Both types of systems are
interesting, and both types offer laboratories for the study of the
rotation-activity relation.

A second reason for undertaking this detailed study is to find very young stars
in the immediate solar neighborhood (i.e., within about 50 pc).  Even if they
were evenly mixed in the Galaxy, very young stars would be rare just because
their ages ($\la100$ Myr) represent such a small fraction of the age of the
Galactic disk.  But such stars are {\it not} evenly mixed because they form in
discrete regions and take time to be dispersed into the field.  There may be a
few stars near the Sun that are as young as, say, the Pleiades, but they are
few indeed.  However, even small numbers matter since they imply that many
more such stars lie in the much vaster volume of the greater solar
neighborhood (i.e., within $\sim100$ pc).  Some of these very young field
stars, such as the HD 98800 system, may be examples of the elusive
post-T Tauri star class; see, e.g., \cite{s98} (1998).

The moderate-resolution spectra obtained for Paper I were centered on the
\ion{Ca}{2} H and K lines to determine the level of CE.  We have obtained
higher-resolution echelle spectra to confirm their high activity levels (at
\ha), to test for youth (with Li), and to observe each star several times to
search for radial velocity changes indicative of close companions.

\section{Observations and Data Analysis}

We obtained high-resolution spectra of our Very Active (VA) sample members
over four nights in 1996 April using the CTIO 1.5-m telescope, the fiber-fed
Bench Mounted Echelle spectrograph, and a Tektronix $2048{\times}2048$ CCD.
We observed 18 stars in 16 systems that were accessible at that time, taken
from Table 6 of Paper I.  This instrument with a 45 ${\mu}$m slit yielded a
dispersion of 0.051 \AA\ pix$^{-1}$ and a measured resolving power (from the
FWHM of Th-Ar lines) of approximately 60,000. A $2{\times}1$ binning across the
dispersion reduced readout time and read noise.  The spectra are centered near
\ha, with coverage from 5600 to 7750 {\AA} after discarding the lowest S/N
orders that are farthest from the center of the CCD.  During the first night
a different cross-disperser was used and the wavelength coverage was somewhat
less: from 5900 to 7110 {\AA}.  Table 1
\marginpar{Tab. 1}
contains a summary of the observations and achieved per-pixel S/N in the \ha\
and Li regions. 

Data reduction used standard IRAF\footnote{IRAF is
distributed by National Optical Astronomy Observatories, which are operated
by the Association of Universities for Research in Astronomy, Inc. under
contract to the National Science Foundation.}
tasks and the specialized routines in the echelle package. Preliminary
processing consisted of overscan subtraction, trimming, and de-biasing.
After flat-fielding with a nightly master flat, the orders were identified and 
traced using the apall package.  Smoothed scattered light corrections were 
made with the apscatter package on the two-dimensional frames.  The resulting
frames were then extracted to one-dimensional orders.  Finally, we applied
dispersion solutions which were determined by fitting the positions of about
600 lines with low-order polynomials in similarly reduced Th-Ar frames; these
solutions had rms residuals of approximately 3 m\AA.

We sought to determine the following from these spectra: a normalized \ha\
emission index, a Li abundance, an estimate of [Fe/H], \vsini, and the radial
velocity (\vrad) and evidence for variability in \vrad\ (Tables 1 and 2).
\marginpar{Tab. 2}
The spectrum features we used are the usual ones for our analyses: \ha\ as
an activity indicator, and \ion{Li}{1} 6708 \AA\ as a test of youth; the
\ion{Na}{1} D lines as an indicator of circumstellar or interstellar material;
lines of \ion{Fe}{1} to determine metallicity; and moderate-strength lines of
Fe and Ca to measure radial velocity.

For \ha, we followed the procedure described in \cite{sshj} (1993b, hereafter
SSHJ).  First we coadded the available spectra and then smoothed them slightly
with a $\sigma = 2$ pixel gaussian.  The \ha\ profile of a star of the same
\bv\ color and known to have little activity was then subtracted to yield the
net emission equivalent width; that was then converted to a flux (\FH) and a
flux ratio (\RH), which is the ratio of the net \ha\ emission flux to the
stellar bolometric flux.  We estimate the uncertainty in \RH\ to be 0.2 dex.
We measured the Li feature both with direct integration and profile fitting.
We corrected for the nearby Fe+CN feature at 6707.4 \AA\ by either fitting a
symmetric profile to the red side of the line, or by using the empirical
correction of \cite{s93a} (1993a).  Both procedures agreed to within a few
milli\AA ngstroms.  $W_{\lambda}({\rm Li})$ was transformed to $A_{\rm Li}
\equiv$ \ALi\ using Table 2 in \cite{s93a} (1993a).  \bv\ colors were used
to determine \teff\ values using equation (3) in SSHJ\@.  The \bv\ colors are
taken from the Tycho catalog for the most part.  For the inactive comparison
stars we have worked with ground-based \bv\ colors, but those differ little
from the Tycho values.  The only significant exceptions are HD 163029NE/SW
for which we used ground-based photometry for reasons explained in the next
section.

We derived Fe abundances for the more slowly rotating VA stars in our sample
which did not show evidence for double lines.  These values were calculated
from the measured equivalent widths of a few \ion{Fe}{1} lines in each
spectrum using an updated version of the MOOG LTE analysis package (\cite{sn73}
1973) and \cite{k92} (1992) model atmospheres.  The atomic data were taken
from \cite{T90} (1990).  Solar abundances were calculated for the same lines
using our sky spectrum and employing the same atomic data.  The errors are
dominated by those in the measurement of the line strengths.  The Fe
abundances for the four stars that could be measured reliably are all within
0.1 dex of solar, though slightly larger excursions are allowed by the
uncertainties.

Rotational velocities for the five apparently-single stars and some of the
probable binaries were estimated from measured breadths of selected metal
lines.  Measurements were not made for the other binary stars because of line
blending.  However, one may easily gauge the qualitative nature of rotation
in these stars from the figures.  Due to their very large projected rotational
velocities, no clean metal features were available to measure \vsini\ for
HD 54759, HD 102982, HD 123732, and HD 175897; upper limits were estimated
from \ha.

We obtained several spectra for most of our targets so that we could detect
short-term changes in the radial velocity.  Stars with periods less than
about 10 days show enhanced CE, and such systems have large radial velocity
amplitudes and usually exhibit obvious changes from one night to the next.
Relative radial velocities from our several spectra were computed, accounting
for both instrumental shifts and heliocentric corrections.  The former were
measured by cross-correlating the telluric B-band region using the fxcor
routine in the IRAF rv suite.  Radial velocities were measured by
cross-correlating the spectra, usually in the 6440-6520 \AA\ region, which 
contains numerous moderately-strong lines.  Because many of these objects
are multiple systems, explanatory notes are needed for the rotational and
radial velocities; these are provided in the next section.

\section{Comments on Individual Stars}

We will first discuss each of the VA objects separately before considering them
in context.  In most cases our observations are the first high-resolution
spectra published for these objects.  Many spectra are composite
so that the strengths of features are diluted by the flux of a companion.  In
other cases the features are extremely broad because of rapid rotation.  Our
primary goal is to distinguish single young stars from close multiples that
show the same high activity levels.

We compare these stars to the Pleiades because many high-quality data are
available for that ZAMS cluster.  Figure 1
\marginpar{Fig. 1}
shows the normalized \ha\ emission flux, \RH = \FH/$F_{\rm bol}$, versus \bv\
for Pleiades stars (SSHJ), shown as small dots.  Stars referred to in this
paper are shown as large circles, with the identifying letters given in
Table 2.  Figure 2
\marginpar{Fig. 2}
shows \ALi\ versus \teff\ for Pleiades stars (\cite{s93a} 1993a) and the
present sample, with the same symbols.  In some cases we also compare the Li
and \ion{Ca}{1} 6717 \AA\ features qualitatively because the strength of the
Li feature generally only exceeds that of the Ca line in very young stars.

\noindent
{\bf HD 37572A} (Star P): This object is both a Rosat and EUVE source
(\cite{p93} 1993; \cite{b94} 1994).  The \ha\ profile (Fig. 3)
\marginpar{Fig. 3}
confirms the very active nature of this object inferred from the Ca H and K
lines in Paper I.  We did not observe an inactive star of exactly the same
color as HD 37572A, but HD 38392 and HD 115617 bracket it in color.  In the
top panel of Figure 3, the similarly-smoothed spectrum of HD 38392 is plotted
as the short dashed line; it can be seen that the HD 37572A \ha\ profile is
filled in compared to the cooler HD 38392.  The resulting difference spectrum
is shown as the long dashed line in the top portion of Figure 3.  The
residual emission was measured by differencing with both standards HD 38392
and HD 115617; the mean equivalent width is listed in Table 2.  The \RH\
values computed with no color adjustment from the two inactive standards,
which differ in \bv\ by a sizable 0.23 mag, differ by about 0.3 dex; thus,
errors in interpolating \RH\ at the VA star's intermediate color should be
small ($\la0.1$ dex).  Comparison with the Pleiades indicates that \RH\ lies
within the distribution of Pleiades stars of similar color.

The bottom panel of Figure 3 displays the coadded three individual spectra of
the Li 6708 \AA\ region; the Li feature is significantly stronger than the
\ion{Ca}{1} 6717 \AA\ line.  Our Li abundance of \ALi=2.78 means that
HD 37572A lies near the upper envelope of Pleiades abundances at
similar \teff\ (Fig. 2), suggesting that HD 37572A is ZAMS or younger.

We checked whether this component of a wide pair ($\sim18$ arcsec separation)
might also be a part of a short-period system.  The two spectra from the first
night have identical radial velocities, and both of these deviate by only 0.6
\kms\ from the spectrum obtained on the last night.  We do not consider the
difference significant.  We find no evidence that the star's activity is due
to membership in a close binary system. 

An absolute radial velocity was measured using the sky spectrum obtained on a
different night.  Instrumental shifts were estimated from the telluric
B-band regions, and radial velocities with respect to the sky were measured
from cross-correlation of the 6500 \AA\ region.  Comparison of the measured
wavelengths of numerous metallic features in the sky spectrum with accurate
laboratory values indicated the need for a 2.7 \kms\ zero-point adjustment.
Applying the zero-point, instrumental, and heliocentric corrections led to
a final mean radial velocity of $+29$ \kms, with an uncertainty of 1 \kms.

In sum, our observations suggest that HD 37572A is a good candidate for a
nearby very young star whose activity and large Li abundance are both
associated with youth rather than membership in a close short-period binary
system. 

\noindent
{\bf HD 41824AB} (Star A): This long-period ($>400$ years) binary system is
both a Rosat and EUVE source (\cite{p93} 1993; \cite{b94} 1994).  \cite{p93}
(1993) suggest that the B component may be a RS CVn system.  Two Coravel
measures, separated by nearly a year, indicate significant radial velocity
variability for the A component, but not the B component (\cite{a85} 1985).  

Hipparcos resolved this system into two stars separated by 2.5 arcsec, but
we did not resolve them, making the spectra appear double-lined.  Figure 4
\marginpar{Fig. 4}
shows two spectra of the Li region acquired on succeeding nights.  Two
components are plainly evident on the first night but not the second.
These spectra have not been adjusted to a rest velocity because of their
double-lined nature and velocity variability.  However, the second night's
spectrum was placed on the velocity scale of the first night by accounting for
a small instrumental ($\sim0.9$ \kms) shift measured from cross-correlating
the telluric B-band regions.  Figure 4 indicates that at least one of the
components is a radial velocity variable on short timescales.  Radial velocity
stability of one of the components like that suggested by the Coravel
observations (\cite{a85} 1985) is the simplest explanation for the
observations in Figure 4, though there is ambiguity in identifying the
specific components in our spectra because of blending.
From the second night's spectrum we determined $\Delta$\vrad=$23.8\pm0.4$
\kms.  The systemic velocity on the first night appears to be $+2.7\pm3$ \kms.

All this suggests that in fact this is a triple system because two stars 30
pc away that are 2.5 arcsec apart will not change their velocities
significantly in one day.  We will call the two components we see ``A'' and
``B,'' with ``A'' being the blueward spectrum in the April 27 spectrum, but
the nomenclature is confused.  The relative line strengths we see are
consistent with the relative magnitudes listed in the Tycho catalog: 7.180
for A and 7.546 for B, and so for the present we will assume that the Tycho
photometry applies to these two spectra and we will label the stars in the
diagrams as ``A1'' and ``A2.''

The second night's spectrum indicates that the rotational velocities of the
two components are not the same because the bluer component has sharper lines.
We estimate \vsini = 4 and 7 \kms, but with large systematic errors of 2 \kms.

The vertical lines in Figure 4 mark the predicted positions of the Li
features.  No Li is evident in either component.  We place a measured upper
limit of 16 m{\AA} on any real feature in the second night's spectrum; this
includes an estimated flux dilution correction of a factor of two.  With the
\teff\ values and abundance methodology as above, we find \ALi$\la1.45$.
Comparison with the Pleiades (Fig. 1 and Fig. 2) indicates that HD 41824 is
not especially young and its activity is likely because one of the components
is a short-period binary.  Additionally, the low resolution spectroscopic 
H and K activity proxy measure from Paper I might be overestimated due to 
effects of flux dilution. 

\noindent
{\bf HD 54579} (Star B): The H and K spectra displayed in Figure 2 of Paper I
showed that this star is broad-lined, and this is confirmed by the \ha\ data.
The \ha\ and Na D lines are the only features clearly seen in our spectra, and
the latter only barely.  The top panel of Figure 5 
\marginpar{Fig. 5}
shows the the first night's coadded \ha\ profile of HD 54579 compared to
HD 81809; the broad-lined nature of the former is readily apparent. Present in
all three of the individual HD 54579 spectra may be an {\it apparent\/} central
reversal marked by the vertical line.  However, this peak does not appear to
be in the center of the \ha\ profile, though the red half of the red wing also
appears to be asymmetric.  We estimated \vsini $\le140$ \kms\ from the \ha\
profile obtained on the last night.  We could not determine a useful \vrad.

The bottom panel of Figure 5 shows the individual spectra from the first night 
and the single spectrum from the second night.  No radial velocity shifts have
been applied, but the alignment of marked telluric features indicates no
significant instrumental shifts.  It can be seen that as the central peak
increases in amplitude during the first night, the width of the blue wing also
increases.  It is possible that the varying blue wing {\it width\/} is in fact 
due to {\it depth\/} variations arising from errors in the continuum
normalization of very broad features in a high-dispersion spectral order of
limited wavelength range; however, the contrary behavior of the central peak
amplitude (depth) and lack of any variations in the red wing are arguments
against this.  The second night's spectrum is narrower, though still very
broad.  Also conspicuous is an absorption ``shelf'' in the far red \ha\ wing.
The centroids of the complex line profiles are, evidently, also different.
While it is difficult to rule out complex emission variability, the simplest
interpretation of the spectra is that there is short-period radial velocity
variability due to a (also broad-lined) companion and that the ``reversal''
noted above is where the two line profiles overlap.

Although our spectra in the D line region are of lower S/N, and have an even
greater uncertainty in continuum rectification due to their lying near the
order edge, the D lines confirm the behavior seen at \ha.  In particular, an
extra blueward component is present in the first night's spectrum, and the
blue wing does move outward in step with the \ha\ wing.  The second night's
spectrum is clearly different, having only a single component (for each D
line) and the profiles having different centroids than those of the first
night.  This provides corroborating evidence that HD 54579 is a short-period
SB2.  As such, its relatively weak HK and \ha\ absorption could also be
affected by flux dilution in addition to activity-related emission.  We note
that there are no published Rosat, Einstein, or EUVE detections for HD 54579. 

\noindent
{\bf HD 102982} (Star C): Recently, \cite{l97} (1997) have found this star to
be both a Rosat and EUVE source.  Our spectrum of HD 102982 indicates the star
is broad-lined, with \ha\ and the D lines being the only clear features.  We
estimated \vsini $\le100$ \kms\ from \ha\ and could not determine a useful
\vrad.  In addition to the Tycho \bv\ value, one can be estimated
using $(b-y)$ from \cite{o93} (1993), \cite{o94} (1994).  \teff=5630 was
inferred from the mean $(b-y)$ using the calibration of \cite{sh85} (1985) and
assuming [Fe/H] = 0.  This \teff\ agrees well with what we inferred from the
Tycho photometry.

The top panel of Figure 6 
\marginpar{Fig. 6}
compares HD 102982 to HD 76151, which has a similar \bv.
The \ha\ line is asymmetric in that the slope of the red wing
appears shallower than that of the blue wing.  This is confirmed by the D
lines, which are shown in the bottom panel of Figure 6.  This behavior is most
simply explained if the star is an SB2, which could also contribute to the
relative weakness of the HK and \ha\ features.  A contact or eclipsing system
might explain the 0.09 mag discrepancy in the $V$ magnitude measured by
\cite{o93} (1993), \cite{o94} (1994) and any genuine emission (as opposed to
flux dilution) inferred from the H and K and Balmer lines.  Unfortunately,
only one spectrum was secured, so we can not address radial velocity
variations.  This star needs further spectroscopic study to draw firm
conclusions; at present, we exclude it from our final young candidate sample. 

\noindent
{\bf HD 106506} (Star D):  The Tycho \bv\ value in Table 2 agrees well with
the 0.58 value we used in Paper I.  However, we estimate \bv $\approx0.65$
from the $(b-y)$ photometry of \cite{o93} (1993), \cite{o94} (1994) and the
color-\teff\ relation of \cite{sh85} (1985) as above.

The top panel of Figure 7 
\marginpar{Fig. 7}
plots the \ha\ spectra obtained on consecutive nights; they differ 
markedly.  The bottom panel indicates that Li is present; moreover, it appears
to be strong compared to \ion{Ca}{1} $\lambda$6717.  The second night's
spectrum seems to have two distinct components.  The strong \ion{Ca}{1} 
$\lambda$6122.2 feature seems to show distinct components on both nights 
(Figure 8).
\marginpar{Fig. 8}
The profile on the second night appears to include a red asymmetry like
the \ha\ feature.  The solid line in the bottom panel of Figure 8 shows
the central region of the cross-correlation function (ccf) resulting from
convolution of this spectrum with our sky spectrum in the 6122 \AA\ region.
The dashed line is the Gaussian fit to the central peak.  The observed ccf is
clearly asymmetric with respect to the Gaussian, or any other symmetric
fitting function one might choose to employ.  Until higher quality data can be
obtained, we classify the star as only a possible SB2.  The D lines may
contain a circumstellar or interstellar component, providing further
justification for higher S/N observations.  While the large Li content of the 
star may be consistent with youth, we do not include the star in our final 
sample of {\it bona fide\/} young candidates since the inferred HK emission 
could be affected by flux dilution or close-companion-related activity. We
also note the lack of Rosat or EUVE detections. 

\noindent
{\bf HD 119022} (Star E): This star was detected as an x-ray source in the
{\it Einstein\/} Slew Survey (\cite{e92} 1992).  \cite{srs96} (1996), however,
suggest that a nearby M star with weak \ha\ emission is more likely to be the
source.  In any case, the HK measurements from Paper I and those alluded to
in \cite{srs96} (1996) suggest significant chromospheric emission for
HD 119022.  The top panel of Figure 9
\marginpar{Fig. 9}
contains the second night's \ha\ profile of HD 119022 and the inactive star
HD 115617.  The broad-lined nature of HD 119022 is readily apparent.  The
bottom panel compares the individual HD 119022 profiles acquired on three
different nights.  The strength of \ha\ varies, but its width does not,
suggesting that HD 119022 is single.  However, Hipparcos has resolved
HD 119022 into two separate objects with a separation of $0.202\pm0.006$
arcsec and $\Delta m = 0.10\pm0.19$ magnitude: There are
two objects of equal brightness contributing to the spectrum.  The radial
velocity in Table 2 was determined from \ha.

The Hipparcos INCA-based color of \bv=0.78 listed in Paper I seems too red for
the G2 spectral type, but it agrees well with the Tycho value listed in
Table 2.  We made an independent \bv\ estimate as before, using
the Str\"omgren $(b-y)$ photometry from \cite{t80} (1980). While this involves
a non-negligible excursion outside the color range of the \cite{sh85} (1985)
relations, a cool \teff\ or red \bv\ is clearly indicated; we find \bv=0.77,
in excellent agreement with the INCA and Tycho values.

Figure 10
\marginpar{Fig. 10} 
indicates that the broad stellar D lines include sharper features.  While
there are numerous telluric lines in the region, we do not believe these
account for the deep sharp features.  In particular, expected telluric
features of comparable or greater strength as those near the position of the
deep sharp features are not visible.  If the lines are of circumstellar or
interstellar origin, their presence could mean that HD 119022 is significantly
reddened.  Higher quality observations would be of interest to establish the
nature of these features, and their implications for reddening. 

The nature of HD 119022 is unclear.  We see no evidence for radial velocity
variations, but the breadth of the lines prevents us from imposing a
stringent limit.  We also cannot confirm any residual \ha\ chromospheric
emission from the second night's spectrum.  Measurements of the HD 119022
\ha\ (absorption) equivalent width are {\it larger\/} than the values measured
in several of the inactive comparison stars that bracket HD 119022 in color.
We do not believe that this circumstance is due to continuum uncertainties or
the presence of telluric features.  On the other hand, we do not have any
reason to doubt the chromospheric emission inferred from the H and K lines in
Paper I and \cite{srs96} (1996).  A possible explanation of the strong Balmer
absorption, the color-spectral type discrepancy, and the narrow features in
the D line profiles may be that HD 119022 suffers from unappreciated and
significant reddening. 

All three spectra of HD 119022 demonstrate the presence of Li, and it is
stronger than the \ion{Ca}{1} $\lambda 6717$ feature.  The presence of
strong Li (we estimate a strength of $\approx 350$ m\AA), the HK emission,
and the broad-lined nature of this star are all consistent with youth, but
it is too luminous to be merely young in the sense used here (on or barely
before the ZAMS) and must instead be a T Tauri-like star (which the
absence of \ha\ emission argues against), or an evolved object.  In any case,
we have not included it in our final sample of the best young candidates.

\noindent
{\bf HD 123732} (V759 Cen; Star F): \cite{heb70} (1970) noted the diffuse and
SB2 nature of this star's spectrum  from objective prism plates.  His follow-up
photometry revealed this star to be a W UMa type binary.  As expected, our
high-resolution spectra indicate the star is both broad-lined and double-lined,
and that it undergoes significant \vrad\ variations on short time scales
(Figure 11).  
\marginpar{Fig. 11}
Our \vsini\ value (Table 2) was determined from \ha.  We exclude  this star
from our final sample of objects whose activity arises from youth.

\noindent
{\bf HD 151770} (Star G): The top panel of Figure 12
\marginpar{Fig. 12}
shows the Li region for this star.  Multiple components are present, and we
classify the star as a spectroscopic triple, and possibly a quadruple.  This
is confirmed from the cross-correlation analysis.  The ccf derived from
convolution with our sky spectrum in the 6480 \AA\ region is shown in the
bottom panel of Figure 12;  three peaks can be seen.  There are intra- and 
inter-night shifts in the line positions and profile morphology, implicating 
some of the components as short-period systems.  Lithium is present in
moderate strength, but it is much too weak to correspond to a pre-main
sequence (PMS) star.  The fact that HD 151770 lies above the main
sequence (see Fig. 25) is probably because it is an evolved system.

\noindent
{\bf HD 155555AB} (V824 Ara; Star H): This well-studied active field star is
known to be
a spectroscopic binary (\cite{bel67} 1967), and is a Rosat and EUVE source
(\cite{p93} 1993; \cite{b94} 1994).  For consistency with other studies, we
have departed from the nomenclature of Paper I; we consider the two
spectroscopic components as A and B, reserving the C designation for the
visual component some 30 arcsec distant.  Historically, HD 155555AB has been
considered a RS CVn system, and is still considered such in some studies.
\cite{p91} (1991), however, suggest that the two spectroscopic components may
be rare examples of post-T Tauri stars.  This was motivated by the very large
Li abundances,  which are confirmed by our observations.  Given its
multiplicity and the persistent uncertainty in this system's evolutionary
status, we do not include it in our final best list of young candidates,
however. 

Nonetheless, two novel results come from our spectroscopy of this star.
First, as demonstrated in Figure 13, 
\marginpar{Fig. 13}
our single spectrum shows \ha\ to be in overt emission, which is different
from the high resolution spectrum in Figure 2b of \cite{p91} (1991).
Therefore, \ha\ appears to be variable over and above any profile changes due
to orbital motion alone.  Second, we find evidence that additional
components may be present in our spectrum.  The bottom panel of Figure 13 
shows the central region of the ccf of the convolution of the HD 155555AB 
spectrum with our sky spectrum in the 6480 \AA\ region.  Two distinct central
peaks are evident.  A broad peak displaced to the red is also seen. The ccf
thus indicates at least three components (but not necessarily individual
stars) are producing the HD 155555AB spectrum.  This is readily apparent from
inspection of the spectrum itself.  As seen in Figure 14, 
\marginpar{Fig. 14} 
there is consistent line-to-line evidence of more than two components.  
Whether the additional components are previously unrecognized  
companions or the manifestation of cool spots on the surface of the two
established components is unclear.  If the profiles' extended red shoulder
-- a consistent line-to-line characteristic -- is due to cool spots, then the
material must have a significant radial velocity relative to the local
photosphere(s).  This is unlikely and we believe this is a triple system.

We determined upper limits to \vsini\ for both A and B of 20 \kms\ by fitting
the peaks of the ccf.  The \vrad\ separation between A and B is 35.5 \kms,
and between A and the red hump in the profiles it is 38 \kms.
 
\noindent
{\bf HD 163029 NE, SW} (Stars J and K): These stars form a visual double with
a separation 2.9 arcsec (Hipparcos), and we were able to obtain separate
spectra for each component.  They are considered Very Active stars in Paper I,
based on a composite spectrum.  No high-energy satellite detections of these
stars have been reported.  The accurate CCD-based \bv\ values of \cite{nsv97}
(1997) for each component represent a substantial improvement over the
uncertain composite color tabulated in Paper I.   However, the true colors of
these stars are uncertain.  The Tycho catalog lists \bv= 0.693 and 0.822 for
SW and NE, respectively, while Nakos gives 0.808 and 1.082 for these two
stars.  The $V$ magnitudes from the two sources also differ.  The wings of
\ha, however, as shown in Figure 16 below, suggest that the Nakos photometry
fits these stars better, and so we use it here.
 
In Figure 15, the spectra of HD 163029NE
\marginpar{Fig. 15}
from the second (top panel) and third (bottom panel) nights are plotted
against the HD 163029SW spectra from both the second and third nights.
Note that the NE component is clearly
double-lined, and both components are easily visible in the Figure. That one
set of these lines does not arise from the nearby SW component is indicated
by the overplotted spectra of the SW component, which (on both nights) are of
clearly different velocity than either of the NE components; we note that 
this does not arise from instrumental shifts.  Thus HD 163029NE is an SB2.

Figure 15 reveals no absolute or differential radial velocity variations of
the NE SB2 system.  This is verified by quantitative cross-correlation
analysis which includes our two spectra from the first night also.  The
maximum radial velocity shifts seen between any two of our NE spectra are
limited to ${\le}0.3$ \kms, which is indistinguishable from zero within the
uncertainties.
 
Figure 15 also indicates that the SW component undergoes radial velocity 
variations on a short (at least day-to-day) timescale.  The velocities on the 
second and third nights differ by ${\sim}40$ \kms; no significant velocity 
variation (${\le}0.3$ \kms) is seen between the two spectra acquired on 
the first night.  While not readily apparent from Figure 15, the lines in 
our spectrum from the third night consistently show shallow absorption
shoulders in the blue wings; these show up most clearly as a notable mild
asymmetry in the cross-correlation peak associated with this exposure.  The
velocity of this component is similar to that of the NE component, so
contamination may be the cause.  However, the possibility remains that SW
is also an SB2.   

Using the \ha\ profile of HD163029 NE to try to detect residual emission is
difficult because of the double lined nature of the spectrum, as well as the
fact that we do not have a standard as red as the NE component (though the
\ha\ profile strength doesn't change so rapidly at these cool \teff\ values).
Thus we rely on the hotter, single-lined component HD 163029SW.  The top panel
of Figure 16 
\marginpar{Fig. 16}
shows the first night's coadded \ha\ region spectrum of HD 163029SW compared
to three stars of different color.  The weakness of the HD 163029SW profile is
apparent, and its residual emission relative to HD 38392 is shown as the
short-dashed line in the bottom panel.  The resulting \ha\ flux is just below
the upper envelope of Pleiades values at similar \bv, implying that it is
similar to the more active Pleiades stars. 

The spectra of the NE component reveal no detectable Li feature.  The Li
abundance level can be reasonably and conservatively gauged by measurement of
a limit to any possible absorption, and combining this with a conservative
factor of two correction for flux dilution.  Our upper limit of
\ALi${\la}+0.35$ at \teff= 4412 is considerably lower than nearly all the
Pleiades stars of similar color or \teff.  Also, we estimate upper limits to
\vsini\ for the blue and red components of 3 and 4 \kms, respectively.
We therefore exclude this object from the young candidate sample. 

All our spectra of the SW component show a weak Li feature.
However, the equivalent width is not well determined.  Our best spectrum gives
a value near 26 m{\AA}, but the other spectra give values from 40 to 55 m{\AA},
yielding a mean value of $41\pm15$ m\AA.  The color-based \teff\ and this
equivalent width lead to \ALi = 1.51.  This value, too, is significantly below
Pleiades values at similar color.  However, it is significantly larger than
Hyades stars of similar color.  This suggests that HD 163029 is an
intermediate-age system, but not especially young.   

In sum, we have conflicting information.  The composite nature of the NE 
component (and possibly the SW too), and the composite (SW+NE) nature of the 
spectrum from Paper I have certainly influenced the estimated strong H and K
emission level reported there.  Also, the uncertain color used in Paper I may
have led to an incorrect estimate of the H and K emission strength.  On the
other hand, we find that the SW component does seem to have weak \ha,
indicating moderate activity.  However, the Li abundances in these stars are
much lower than those seen in the Pleiades and other young cluster and field
stars of similar age, which could suggest an intermediate age.  While the Li
constraints are not stringent and might be disregarded (though remain
interesting), the SW component is a spectroscopic binary; thus, this might
conceivably contribute to the origin of its chromospheric emission rather than
youth.  Because of this conflicting information, we exclude the system from
our list of young candidates.  

Because radial velocities are lacking for the HD 163029 stars, we derived
values from cross-correlation of the 6440-6520 {\AA} region using our sky
spectrum.  Instrumental shifts and zero-point corrections were made as
described for HD 37572.  We find a heliocentric radial velocity of $-28.0$
\kms\ for HD 163029 NE.  The uncertainty, given the double-line nature of the
spectrum and the irregularly shaped correlation peak, is perhaps 3 \kms.
Given the marked variations and the small number of spectra, a systemic
measure for SW will also be uncertain. The mean night-1 velocity, which is the
median heliocentric systemic estimate, is $-26.2$ \kms.  This agrees well with
the NE component; however, there are no individual proper motions to confirm a 
physical association.     

\noindent
{\bf HD 174429} (PZ Tel; Star Q): HD 174429 is a well-studied VA star from
Paper I
known to demonstrate HK emission since the work of \cite{bm73} (1973).
HD 174429 is a Rosat XUV source (\cite{kbs95} 1995) and a stellar radio source
(e.g., \cite{lw95} 1995).  The \ha\ profile is known to be filled in 
(\cite{ict88} 1988).  No spectroscopic evidence of duplicity has ever been
noted, and most recent studies consider this object to be single, with a
rotation period of 0.94 days (\cite{ict88} 1988).  \cite{ict86} (1986)
suggested that the star was a member of the local Pleiades group from its
kinematics, its very strong Li feature, and its significant rotation (rapidly
rotating cool Pleiades stars also demonstrate very large Li abundances).
Combining \vsini\ and rotational period measures, \cite{rgp93} (1993) note the
implied radius is too large for a main-sequence star, but indicative of PMS
status.  With respect to the PMS versus RS CVn classification, HD 174429 is
similar to HD 155555AB\@.  Perhaps additional PMS field candidates could be
found by rigorous reinvestigation of systems classified as RS CVn.  

Our \ha\ profile confirms the Balmer line weakness.  The top panel of 
Figure 17
\marginpar{Fig. 17} 
contains the \ha\ region of HD 174429 and the cooler inactive standard
HD 38392; the weakness of the former's Balmer line is clear.  Values of the
\ha\ emission flux, listed in Table 2 and computed using the standards
HD 38392 and 115617, place the emission in the upper portion of the Pleiades
distribution at similar color.  As previous spectroscopy has shown, the Li
line is very strong -- exceeding the strength of the \ion{Ca}{1} 
${\lambda}6717$ line (bottom panel of Figure 17).  Our measured equivalent 
width, corrected for the \ion{Fe}{1} 6707.4 \AA\ contribution as before, is
smaller than that measured by \cite{prg92} (1992), but is within the
uncertainties.  The \bv\ photometry from \cite{cl97} (1997) is in good
agreement with the Tycho value used here. Our LTE abundance of \ALi=3.11 is
some 0.8 dex lower than the value of Pallavicini \etal\ value of 3.9; the
difference seems larger than can be accounted for by small or modest \teff\
and line strength differences, but differences in other aspects of the
analyses (model atmospheres, damping, etc.) may be non-negligible.  Comparison
with the Pleiades shows that the HD 174429 Li abundance is equivalent to that
demonstrated by the young cluster's rapid rotators of similar color. 

A radial velocity determination was made by cross-correlation of three
different orders with our sky spectrum.  Combining the instrumental shift and
zero-point correction gives a heliocentric velocity of $-13.5$ \kms.  Possible
systematic errors may be as large as 2.5 \kms, and total uncertainties are
near 3 \kms.  Our velocity is significantly smaller than the $+4.4{\pm}6.2$
\kms\ estimate of \cite{b87} (1987), and the $-3.2{\pm}3.7$ \kms\
estimate by \cite{ict88} (1988).  Nevertheless, the $V$ velocity indicates
that HD 174429 is {\it not\/} a member of \cite{e75} (1975) Local Association
(Pleiades group), which has a well-defined value of $V=-25$.
 
Given the HK and \ha\ emission, the large Li abundance, and the
lack of any convincing evidence for duplicity, we include HD 174429 as a young 
candidate. 

\noindent
{\bf HD 175897} (Star R): This star is a far-UV point source
(\cite{bs95} 1995).  The \ha\ profile from the third night's spectrum is
plotted with the profile of the similar color standard HD 76151 in Figure 18. 
\marginpar{Fig. 18}
The broad-lined nature of the VA star is clear.  \ha, the D lines, and the
\ion{Fe}{1} ${\lambda}6136.6$, 6137.7, 6141.7 blends are the only obvious
stellar features.  There may be marginal evidence for small changes in the
\ha\ emission strength, the possible appearance of a small central emission
reversal on the first two nights, and a possible asymmetry in the blue wing
relative to the first two nights.

Our measurements of \ha\ seem to confirm the VA classification, but given the
uncertain \bv\ color from Paper I, we first derived another estimate from the
$(b-y)$ photometry of \cite{o94} (1994) as described before.  We find \bv=0.68,
somewhat redder than the Paper I value of 0.61, the Tycho value of 0.65, or
the spectral type of G0.  The \ha\ equivalent width was measured for HD 175897
and the similar color standards HD 76151 and 81809.  Interpolating among the
small color differences we find the residual emission and \ha\ fluxes listed
in Table 2. The \ha\ emission is in the middle of the Pleiades distribution. 

Radial velocities were derived from cross-correlation of the \ha\ region.
While not ideal, this is the only feature with enough statistical power to
derive relatively secure results.  We find no evidence for night-to-night
variations larger than the apparent uncertainties.  There may be asymmetries
in the cross-correlation peak which result from the oddities noted above, but
the significance is difficult to assess.  Our heliocentric velocity estimate
is $+20.7$ \kms, with an estimated uncertainty of 4.5 \kms, but it is
possible that the \ha-based velocity does not precisely correspond 
to the systemic stellar velocity. 

Given the VA nature of the star as determined from Paper I, our confirmation 
of this from \ha, and the lack of any evidence of duplicity, we have included 
HD 175897 as a young candidate.  Additional monitoring and higher S/N spectra 
would be welcomed to investigate the possible spectral oddities noted above, 
as well as to derive a more secure limit to or value of the Li abundance. 

\noindent
{\bf HD 177996} (Star L): This object has no high energy satellite detections.
Spectra in the \ha\ and Li regions are shown in Figure 19;
\marginpar{Fig. 19}
the star  is clearly an SB2.  While there are only small (a few \kms)
radial velocity variations in the A component, large and significant shifts
are present for the secondary; the period must be short (i.e., a few days or
less).  On the first night, the spectral lines of the two components coincided
at a \vrad\ of $-38.4\pm1.5$ \kms.  Relative \vrad\ values on the other
nights are noted in Table 2.

While both objects are not far from being sharp-lined (as defined by our
instrumental resolution), the secondary appears to have broader lines than  
the primary.  Li is present in the primary and the secondary 
also.  Simple comparison relative to the $\lambda$6717 \ion{Ca}{1} line 
indicates Li is much lower than (unevolved) stars of similar color in young 
clusters.  A rough estimate of the flux-corrected Li equivalent width of the 
primary is ${\sim}25$ m{\AA}, which we believe accurate to 5 to 10 m{\AA}.  
Although we adopt the color and \teff\ of the (now known to be) composite
system, which is presumably systematically redder/lower than the A component
alone, this is not a dangerous assumption because the abundance sensitivity
to adopted \teff\ is not expected to be any steeper than the observed
\ALi-\teff\ morphology of young and intermediate-age cluster stars at the
approximate \teff.  We derive log $N$(Li)${\sim}1.10$, which is significantly
less than single stars or short-period binaries of similar color in the
Pleiades.  However, it is significantly larger than that of single Hyades
stars, and comparable to cooler short-period Hyades binaries.

Given the SB2 nature of the star (which may have diluted the H and K
absorption), the moderately low Li abundance, and lack of x-ray or EUV
detections, we do not consider the star as a young candidate. 

\noindent 
{\bf HD 180445} (Star M): Our spectra are some of the few observations of any
sort for this poorly-studied VA star; it has no satellite detections.  The
\ha\ spectra confirm the HK-based VA designation.  The Balmer line is very
weak -- nearly filled in on the first night of observations -- and variable in
strength.  This weakness is not due to dilution from a companion, though the
second and third night's data have a weak second set of lines.  

The top panel of Figure 20 
\marginpar{Fig. 20} 
contains the \ha\ spectra from all three nights.  Nightly velocity shifts, 
significantly larger than the tiny instrumental and heliocentric shifts, are 
evident.  The period of this SB must be short (days or less).  The bottom 
panel compares the 6410 \AA\ regions of the Sun (dashed line) and the 
second night's spectrum of HD 180445 (solid line).  A weak second set of lines 
appears in the stellar spectrum, and is visible in other orders too.  The 
second set is less apparent on the first night, and not identifiable on the 
third night.  Part of these changes could be related to lower S/N in some of
the spectra from the second and third nights.

The \vrad\ of the A component on the first night is $-26.2\pm0.6$ \kms, but
the systematic error may be as large as 15 \kms\ due to effects of the
secondary.  The second night's spectrum yields a \vrad\ difference of
$-105\pm7$ \kms\ (A minus B), and on the third night this difference is
$+85\pm5$.

No consistent feature ascribed to Li can be seen in our three spectra.  We are
able to place a conservative limit (including any flux dilution) of 50 m\AA\
on the equivalent width of any Li line.  This is about a factor of two less
than the line strengths of Pleiads of similar color.  While the resulting
abundance upper limit is  significantly larger than Hyades or Praesepe stars
of similar color, the low Li abundance upper limit is considerably smaller
than that seen in very young stars of any metallicity (Fig. 2).  Given the
binary nature and the low Li abundance, we do not include HD 180445 in our
young candidate sample. Its \ha\ does appear to be genuinely filled in with
emission, however, and this is probably due to rapid rotation induced by a
close, tidally-locked secondary.

\noindent
{\bf HD 202917} (Star S): This VA star, otherwise not well-studied, has
recently been classified as a Rosat and EUVE source (\cite{l97} 1997).  The
top panel of Figure 21 
\marginpar{Fig. 21}
plots the \ha\ profile of HD 202917 with that of the slightly cooler
standard HD 115617.  The VA classification of Paper I is confirmed by the 
weakness of the Balmer line.  We measured the residual \ha\ emission as before 
using the standards HD 115617 and 76151, which bracket HD 202917 in color.  The 
measurements, listed in Table 2, indicate that the \ha\ flux of HD 202917 lies 
near the upper envelope of the Pleiades distribution for its color (Fig. 1).

The bottom panel of Figure 21 shows the first night's coadded spectra in the 
Li region.  The Li line is very strong, surpassing the \ion{Ca}{1}
$\lambda$6717 feature in strength -- again, a rare occurrence generally
limited to young stars.  The measured Li line strength was corrected for
\ion{Fe}{1} $\lambda$6707.4 contamination, and the resulting equivalent width
and (LTE) abundance are listed in Table 2.  The abundance is a few tenths of a
dex larger than that in Pleiads of similar color (Fig. 2), despite the fact
that HD 202917 does not appear to be a rapid rotator; we estimate
\vsini$\approx$ 10-15 \kms.

The radial velocities from the first night's two spectra are indistinguishable.
We find no radial velocity difference between the first and second night
larger than 0.35 \kms, which is within the errors.  \cite{e86} (1986) lists
the star as having a variable radial velocity, based on a private communication
of unpublished work.  Without more detail, it is impossible to say if our
(quite limited) data conflict with such an assessment.  We initially wondered
if the first night's spectra might contain asymmetric profiles indicative of
an SB2 classification.  Given the modest S/N, we coadded the spectra and
cross-correlated the data with our sky spectrum in the 6480 \AA\ region.  The
cross-correlation function appears symmetric, a conclusion that held whether
utilizing filtering or not.  We measured a radial velocity from our spectra,
with a zero-point and instrumental correction estimated as before.  Our
heliocentric value of $-5.3$ \kms\ is in good agreement with the $-7$ \kms\
value from \cite{e86} (1986); the uncertainty in our velocity is near 1 \kms. 

Given the significant HK and \ha\ emission, large Li abundance, x-ray and EUV
detections, we include HD 202917 as a young candidate lacking convincing
evidence for duplicity. 

\noindent
{\bf HD 222259 A, B} (Stars T and N): This VA system from Paper I is both a
Rosat and EUVE source (\cite{b94} 1994).  We observed the components of this 5
arcsec visual double individually.  Before proceeding with analysis of our
spectra, we remedy the prior lack of individual \bv\ colors or an accurate
composite color for these stars by adopting the photometry of the Hipparcos
Double and Multiple component solutions.  These new values are listed in 
Table 2; the uncertainties are a few hundredths of a magnitude.

The bottom panel of Figure 22
\marginpar{Fig. 22}
shows the individual \ha\ profiles of the B component compared to the slightly
warmer standard HD 38392.  The overt \ha\ emission in the first spectrum
confirms the VA nature of this star despite any flux dilution from a
companion(s).  The \ha\ absorption seen in the spectra acquired only 16
minutes later as well as on the next night suggests significant \ha\
variability on short timescales.  Small but significant radial velocity shifts
between the spectra are observed.  Moreover, the line profiles also change,
becoming significantly narrower in our second spectrum.  The radial velocity
variations indicate a SB designation; while no clear second set of lines is
identifiable, the profile morphology changes indicate an SB2 designation.
Apparently, this is a close system, with a period of $\la2$ days. 

The Li features in all the spectra of both the A and B components are 
comparable in strength to the \ion{Ca}{1} $\lambda$6717 feature (Figure 23), 
\marginpar{Fig. 23}
indicating relatively large Li abundances.  Estimates for both the A and B
components were derived as before. For the binary B component, the mean
measured line strength (abundance) from the three individual spectra might be
overestimated by a factor perhaps as large as 2 (0.5 dex) due to blending,
though this is likely quite a generous bound.  An underestimate due to flux
dilution is also possible, but would only strengthen the conclusion of a large
Li abundance.  The raw equivalent widths were corrected for \ion{Fe}{1}
$\lambda$6707.4 contamination, and the derived LTE abundances are listed in
Table 2.  The value for A is large -- comparable or a couple tenths of a dex
greater than the majority of Pleiades values at similar color and projected
rotational velocity (Fig. 2).  The B component abundance is also large, and
appears to fall on or perhaps slightly above the upper envelope of Pleiades
abundances, which show significant scatter.  

The HK emission from Paper I, the Balmer line emission measured here,
the extant x-ray and EUV satellite detections, the lack of convincing evidence
for duplicity, and the very large Li abundance all make HD 222259A a promising
young candidate system.  This also means that HD 222259B should be a promising
young candidate system too, but given our close binary criterion, we have not
labeled it as such in Table 2.  The B component, then, is a specific example
of a binary system that is most likely a young object, but which is rejected
by our exclusive criteria.  As mentioned before, there may be other young
binaries we have similarly excluded.  

Absolute radial velocities for the A and B components were measured from
cross-correlation analysis of the 6440-6520 {\AA} region as before.  For the
variable B component, the first spectrum (having the median velocity) was
taken to be the best statistical estimate of the systemic velocity.  The
resulting heliocentric values are reported in Table 3 (see below).  The
uncertainties are estimated to be near or slightly below ${\sim}1$ \kms;
systematic errors could be significantly larger for the B component of course,
but there is good agreement between the A and B values.  During the analysis we
noticed the cross-correlation functions of the A component were slightly
asymmetric, more so than the B component functions despite the lack of
evidence for double lines.  The asymmetries (an extended blue tail) result in 
increased sensitivity of our derived radial velocities to the fitting function,
and such uncertainties are the dominant item in our error budget.  The
asymmetries are not present in the cross-correlation function of the
telluric B-band features, suggesting the result is not instrumental.  That the
cross-correlation tail is preferentially extended in the same direction for
both A and B indicates that the cause is not due to contamination of the given 
component's spectrum by the other component.  Additional observations of
higher S/N and resolution would be of interest to determine if the cause is
an additional heavily blended set of weak lines in the A component. 

\section{Discussion}

Of the 18 stars in 16 Very Active systems from Paper I that we have observed
at high resolution, 13 show evidence of being close binaries on the
basis of either their line profiles or variable radial velocity.  It is also
possible, of course, that some of our young candidates are members of close
binary systems that we did not recognize.   While this close binary fraction,
nearly 75\%, is high, it is not surprising given that our sample was composed
of stars with strong chromospheric emission.  In Figure 24,
\marginpar{Fig. 24}
we reproduce Figure 7(a) of Paper I, showing the distribution of chromospheric
emission strengths of the 600+ stars observed there.  We have updated the
upper portion to reflect what we now know about duplicity in these systems.
We have indicated known or probable doubles with crosses.  In this updated
Figure we have added the three targets from the secondary sample of Paper I
and have adjusted \lR\ values using the newer \bv\ values of Table 2.

Five of our eighteen VA stars are probably young stars (six if HD 222259B
is included).  The \ha\ emission levels and lithium abundances of these stars
(see Fig. 2) confirm their youth.  This number may be an underestimate since we
may have excluded some young stars because they were in close binaries.  In
particular, we noted that HD 106506 may be a young SB2 and that HD 155555AB
has been suggested to be a post-T Tauri star by others.  Also, HD 119022 could
be a young, heavily-reddened star.  In this respect, although excluded from
our final young candidate list, HD 119022 could be one of the more interesting
objects in our sample because of its high luminosity.  Clearly, these possible
young stars merit further study.  

For the five apparently single young candidates, our \ha\ measurements
confirm the VA assignment made in Paper I.  Indeed, three of the five stars
are at or above the upper bound of the Pleiades \RH--\bv\ distribution for
stars of their color.  With the obvious exception of HD 119022, the apparent
weakness of \ha\ absorption in the remaining VA objects -- including the overt
\ha\ emission for HD 155555AB and HD 222259B -- is fully consistent with the
weak \ion{Ca}{2} absorption identified in Paper I\@.  In this sense the
results of the low resolution \ion{Ca}{2} spectra in Paper I are reliable.
The present work, however, indicates that in a significant number of cases
determining whether those spectra reliably measure the effects of chromospheric
emission (due either to youth or a close companion) or the effects of flux
dilution from a companion (or some combination) may require follow-up
monitoring with high resolution spectroscopy.  Also, some of the \bv\ colors
used in Paper I were very uncertain, leading to possibly spurious
identifications as Very Active systems, but that appears to have been rare.

Figure 25
\marginpar{Fig. 25}
shows a color-magnitude diagram for 18 stars in our sample for which Hipparcos
parallaxes (\cite{esa97} 1997) are available (none was available for
HD 102982).  Three of the probable single stars (shown as circles) appear to
be PMS objects, with ages of about 10 to 15 Myr, putting them in the
post-T Tauri class of stars.  The other three have positions in the CMD
consistent either with ZAMS or PMS evolutionary status.  The multiple systems
(shown as squares in Fig. 25), are most likely RS CVn systems which lie above
the main sequence because they are evolved.

In Table 3, we present the kinematics of our most-likely young stars, using
\marginpar{Tab. 3}
the radial velocities derived here.  The proper motions are from the PPM
catalog (\cite{br93} 1993), and the parallaxes are from Hipparcos.  The $UVW$
space motions were calculated with an updated version of the prescription
from \cite{js87} (1987).  

There are several noteworthy kinematic properties of our final young 
candidates.  First, the $UVW$ velocities of HD 175897 are in excellent
agreement with those of the Pleiades cluster (\cite{e83} 1983).  This
is not conclusive evidence that these stars in fact have their origin in the
Pleiades because that part of velocity space is near the Local Standard of
Rest, where young stars are likely to be found in any case.  If these stars
indeed form a Pleiades halo, then the volume of space we have surveyed
compared to the distance to the Pleiades suggests that halo may contain a
hundred or more stars, especially since their density is likely to be higher
in the immediate vicinity of the cluster itself. HD 37572A may be another star
associated with the Pleiades. 

We previously noted that our assessment of the kinematics of HD 174429 does
not place it in the Pleiades group, as had been suggested by \cite{ict86}
(1986).  Our radial velocity disagrees with theirs.  Also, the Hipparcos-based
distance is some 25\% smaller than their assumed value.  The origin of this VA
young candidate star, which has a very large Li abundance, is unclear. 

We note that the $U$ and $V$ velocities of HD 202917 and HD 222259A agree
well with each other.  Furthermore, these velocities also agree with the
values for HD 98800, which \cite{s98} (1998) argue is a rare example of a
post-T Tauri star.  The small differences in the $U$ velocities (2 to 3 \kms)
between HD 98800 and the other two stars are consistent with their spatial
separations (about 60 pc) given their ages ($\sim10-20$ Myr).  The  two VA
young candidates and HD 98800 also share the property that their Li abundances
are comparable to or larger than the Pleiades stars of similar color and
projected rotational velocity.  Thus, we suggest that HD 202917 and 222259 may
be two more examples of post-T Tauri stars.  The ages for these stars from
Figure 25 are about 20 Myr or more, older than the $\sim10$ Myr age of the
HD 98800 system (\cite{s98} 1998), but perhaps consistent to within the
uncertainties.  Furthermore, these three systems may provide evidence of a
small group of very young stars sharing a common origin, supporting the
suggestion of \cite{k97} (1997) that HD 98800 is accompanied by several other
PMS objects.  If this interpretation is correct, then HD 222259 would be
another example of a multiple post-T Tauri system that could provide
constraints on the origin of post-T Tauri stars in the field (\cite{s98} 1998). 

An occurrence of five very young stars means that $\sim1$\% of the total
sample of Paper I is younger than the Pleiades.  Since the Pleiades is 100
Myr old and these stars have lifetimes of $\sim10$ Gyr, this fraction seems
reasonable.  However, 1\% is, in fact, high.  First, very old stars are
missing from the solar neighborhood because of disk heating, meaning that
five stars corresponds to $\sim1.5$\% once this effect is compensated for, but
this is unimportant given the numbers of stars involved.  More seriously, we
expect few, if any, stars to have found their way from star-forming regions
into the immediate solar vicinity in such a short time.  The Pleiades itself
could supply some of these stars because a star need only move at $\sim1$
\kms\ relative to the cluster to cross the $\sim100$ pc from there to here in
100 Myr.  But these very young stars in fact appear to be much younger than
the Pleiades and do not have the kinematics of that cluster.  Getting a star
from, say, Taurus-Auriga to the solar neighborhood in 10 Myr requires a
peculiar velocity of 15 \kms\ or more, which is substantial.

\section{Li Abundances in the \ha\ Standards}

Since they may be of interest, we have derived Li abundances for our inactive
\ha\ standard stars.  We expect these stars to be old and to
have low Li abundances.  Our results, given in Table 2, confirm this.
Detailed spectrum synthesis could probably provide tighter constraints on the
abundances, but we have chosen, as a reliable expedient, to simply measure
equivalent widths in the spirit of our cursory examination of these stars'
abundances.  The results may provide a starting point for studies of very low
Li abundances in older solar-type stars. 

Peripheral notes about two stars may be of interest.  First, Li is detected 
in HD 76151; despite being of solar \teff\ or slightly cooler, the Li 
abundance is some four times larger than the Sun's.  (We also detect Li in
HD 38393 and HD 45067, but this is not unusual at their relatively early
spectral types.) Second, because we lack precision multi-band photometry for
both components, we have assumed HD 158614N and S to be identical although
orbital determinations and older photometry indicate a slight difference.  Our
spectrum of each component is contaminated by the other.  There is some
indication in our spectra that Li may be present in HD 158614N.  Extremely
high resolution (spatial and spectral) and S/N spectra, and precision resolved
photometry would be worthwhile since a difference in the components' Li
abundance might be of interest. 

\section{Conclusions} 

We present high resolution echelle spectroscopy of 18 Very Active southern
solar-type stars identified in the \ion{Ca}{2} H and K survey of \cite{h96}
(1996).  We find evidence from line doubling or radial velocity variations
that 13 of these are members of short-period, close binary systems.  Activity
in these stars may thus be due to the presence of a close companion, rather
than youth; however, it is entirely possible that some of the objects
(HD 106506, 119022, 155555AB, and HD 222259B) are young close binaries.  Just
a few exposures of high resolution but very modest S/N seems to be a highly
effective and efficient means of identifying active close binary systems.  

Based on our \ha\ observations, we confirm that the remaining five of the
eighteen stars are also Very Active, and find no evidence that such activity
is due to membership in a close binary system.  Four of these stars also have
significant Li abundances, comparable to or larger than Pleiades stars of
similar color.  We consider these five stars to be young candidates, i.e.,
stars whose chromospheric activity seems associated solely with youth.   

We note that two of the young candidates (HD 202917 and HD 222259) appear to
have the same $U,V$ velocities as HD 98800, a rare example of a post-T Tauri
star in the field, according to \cite{s98} (1998).  While the formation site
of these stars is unclear, the three stellar systems may be part of a small
group of very young stars sharing a common history and origin.  The $UVW$
velocities of HD 175897 are in excellent agreement with those of the Pleiades.

Two other interesting objects are HD 106506 and HD 119022, which were noted
above to be possible young objects excluded from our final best young candidate
list due to their duplicity.  Both of these objects demonstrate sharp features
in the broader Na D lines and stronger features of other metals.  We do not
observe any radial velocity shifts of these sharp features.  This would
suggest an interstellar or circumstellar origin for the sharp D-line features.
These two objects are among the most distant in our sample: the
Hipparcos-based distances are ${\sim}125$ pc, but seems too near for an
interstellar origin.  There is thus the intriguing possibility that these
features arise from circumstellar material, perhaps not unlike that
surrounding ${\beta}$ Pictoris or early-type shell stars similarly inferred
from the presence of sharp features in a broader stellar absorption line.
Comparison of the spectral type and photometry for HD 119022 suggests a
moderate-sized reddening of perhaps 0.1 to 0.15 mag in $E(B-V)$.  Higher
quality spectra and high resolution imaging of these objects would be of great
interest. 

Finally, Li abundances or upper limits were derived for the sharp-lined
inactive \ha\ standards employed in our program.  We report detectable Li for
the solar \teff\ star HD 76151 and for the late F-stars HD 38393 and 45067.
There may be a Li abundance difference in the two similar wide components of
HD 158614; better knowledge of their fundamental parameters from spatially
resolved photometry and spectroscopy is needed. 

\acknowledgments
This work was supported in part by NASA grant NAGW-3695.

\newpage

\figcaption{The normalized \ha\ emission flux, \RH, versus \bv\ for the
Pleiades (small dots, from SSHJ) and the present sample (large circles).  The
letters inside the circles identify the stars; see Table 2.  (The two vertical
dashed lines indicate the range of possible \RH\ values in two Pleiads; see
SSHJ.)}

\figcaption{Lithium abundance (\ALi, on a scale where $\log N({\rm H}) = 12$)
versus \teff\ for the Pleiades (small dots, from \cite{s93a} 1993a) and the
present sample (large circles and triangles).  The letters inside the symbols
identify the stars; see Table 2.  Triangles indicate upper limits.}

\figcaption{In the top panel, the \ha\ region spectra of HD 37572A (solid line) 
and the HK standard HD 38392 (short dashed line) are overplotted.  The residual 
difference spectrum renormalized to continuum level is shown as the long
dashed line.  The bottom panel displays the coadded spectrum of HD 37572A
in the 6707.8 \AA\ Li region.}

\figcaption{Spectra of HD 41824AB acquired on consecutive nights are
overplotted.  Evidently, at least one of the components is a radial velocity
variable.  The bluer component in the second night's spectrum has sharper
lines.  There is no clear Li feature at the expected positions (vertical
lines).} 

\figcaption{The top panel shows the summed \ha\ profile of the VA star HD 54579 
(solid line) and the similar \bv\ color standard HD 81809 (dashed line).  Both 
spectra have been mildly smoothed by convolution with a Gaussian kernel having
${\sigma}=1$ pixel.  The broad-lined nature of the VA star is apparent.  The
vertical line marks a possible central reversal.  The bottom panel displays
individual \ha\ profiles of HD 54579 which have been more heavily smoothed
(${\sigma}=3$ pix) to reduce clutter.  Telluric features are marked with a
${\oplus}$.  A progression in the extent of the blue wing and the candidate
central reversal amplitude in the first three spectra is apparent.  The second
night's spectrum is markedly different.} 

\figcaption{The top panel shows the \ha\ profiles of the VA star HD 102982 
(solid line) and the similar \bv\ color standard HD 76151 (dashed line). 
Spectra have been lightly smoothed with a ${\sigma}=1$ pix Gaussian. The bottom 
panel shows the smoothed Na D features in HD 102982.  Both the D
lines and Balmer line show a red wing shallower than the blue wing.}

\figcaption{The top panel plots the \ha\ profiles of the VA star HD 106506.  
The spectra have been smoothed as above.  No absolute or velocity shifts have
been applied; any relative instrumental and heliocentric shifts are negligible.
The bottom panel shows the Li $\lambda$6707.8 and \ion{Ca}{1}
$\lambda$6717 regions.} 

\figcaption{The top panel shows the lightly smoothed ${\lambda}6122$ 
\ion{Ca}{1} profiles of HD 106506.  The bottom panel shows the central 
region of the cross-correlation function (solid line) resulting from 
convolution with the sky spectrum in the 6122 \AA\ region.  The 
dashed line is a Gaussian fit to the peak of the ccf.  The horizontal bar
in the top panel corresponds to the same range of velocity encompassed
by the cross-correlation in the lower panel.}

\figcaption{In the top panel, the smoothed second night \ha\ profiles of the 
VA star HD 119022 and the standard HD 115617 are compared to illustrate the
broad-lined nature of the former.  In the bottom panel, the smoothed HD 119022
profiles from the individual nights are compared.  Two regions severely
afflicted by particle events have been excised from the first night's
spectrum.}

\figcaption{The lightly smoothed second night's spectrum of the Na 
D region in HD 119022.} 

\figcaption{The unsmoothed spectra of the \ha\ region of HD 123732 
acquired on consecutive nights shows the broad- and double-lined nature 
of this W UMa system.} 

\figcaption{The top panel shows the mildly smoothed spectra of HD 151770 in 
the Li region from the first and second nights of observation; the spectra 
are offset in flux by an additive constant.  Multiple peaks are seen in the 
\ion{Ca}{1} ${\lambda}6717$, Li, and \ion{Fe}{1} ${\lambda}6703, 6705$
features, all of which are blueshifted by ${\sim}1.5$ {\AA} from rest
wavelength.  The different components are best seen in the \ion{Ca}{1}
$\lambda6717$
feature.  The bottom panel shows these three peaks in the cross-correlation
function of HD 151770 with a sky spectrum in the 6480 \AA\ region.
The three vertical bars in both panels are located at the same relative
velocities, to show how the cross-correlation confirms the line structure
seen in the 6717 \AA\ feature.}

\figcaption{The top panel contains the mildly smoothed \ha\ spectrum of
HD 155555AB, which shows a Balmer line in overt emission.  The bottom
panel contains the central region of the cross-correlation function from
convolution with the solar spectrum in the 6480 \AA\ region.  Two
distinct central peaks are seen in addition to a broad 
one to the red; this behavior can be seen in the spectra themselves (Fig 14).
The vertical and horizontal bars in the lower panel correspond in velocity
to the same markings in Figure 14.}

\figcaption{Mildly smoothed spectra of HD 155555AB in the \ion{Fe}{1}
${\lambda}6408.0, 6411.7$ (top panel) and Li (bottom) regions.  The A and B
components are marked in the top panel.  The pronounced red shoulder,
consistent from line-to-line, is marked with a question mark.  Another
possible component between the A and C components is also marked with
a question mark.  The comparison star is HD 76151.}

\figcaption{In the top panel, the mildly smoothed spectrum of HD 163029 NE
(solid line) from the second night is plotted with spectra of HD 163029 SW
(dashed lines) from the second and third nights.  In the bottom panel the
spectrum of HD 163029 NE from the third night is plotted against the same SW
spectra.  The vertical lines mark the two components of HD 163029 NE.}    

\figcaption{The mildly smoothed \ha\ profile of HD 163029SW from the first 
night (solid line) is plotted with the VA star HD 37572 (long dashed line), 
itself inferred to have weak \ha.  The residual emission of HD 163029SW 
is shown as the short dashed line.} 

\figcaption{The top panel contains the mildly smoothed \ha\ profiles of 
HD 174429 (solid) line and the cooler inactive standard HD 38392 (dashed 
line). The bottom panel shows the smoothed $\lambda$6707.8 Li
and $\lambda$6717 \ion{Ca}{1} profiles of HD 174429.}

\figcaption{The mildly smoothed \ha\ profile of HD 175897 from the third 
night (solid line) is plotted with the the \ha\ profile of the
similar-color standard HD 76151 (dashed line).  The very broad-lined nature of 
the VA star is clear.} 

\figcaption{Mildly smoothed spectrum of HD 177996 from the second night of 
observations; no radial velocity corrections have been applied.  The top 
panel contains the Li region.  The two components of the (rest wavelength) 
\ion{Fe}{1} $\lambda$6703.6, $\lambda$6705.1, $\lambda$6710.3, and \ion{Ca}{1} 
$\lambda$6717 features are marked.  The bottom panel shows the \ha\ region 
where, again, two sets of lines are evident.}

\figcaption{The top panel shows mildly smoothed \ha\ profiles of HD 180445 from 
all three nights.  There are clear variations in radial velocity and the 
very weak Balmer absorption.  The bottom panel displays spectra of the daytime
sky (dashed line) and HD 180445 (solid line) in the 6410 \AA\ region.
A second set of lines, seen in the other orders too, is evident in the stellar
spectrum.}

\figcaption{The top panel shows the first night's coadded and mildly smoothed 
\ha\ spectrum of HD 202917 (solid line), compared to the slightly redder and
cooler standard HD 115617 (dashed line).  The mild macroscopic broadening
and weakened \ha\ for the VA star is apparent.  The bottom panel shows the
coadded and mildly smoothed spectrum of the Li region in HD 202917.} 

\figcaption{The top panel shows the co-added and mildly smoothed \ha\ profile 
of HD 222259A (solid line) compared to the profile of the standard HD 81809,
which is of nearly identical color; the modest macroscopic broadening and a
significantly weakened Balmer line of the VA star are evident.  The bottom
panel shows \ha\ profiles of HD 222259B from the individual spectra (lines)
compared to the slightly hotter standard HD 38392 (dots).  The spectra have
been offset vertically for clarity.  Overt \ha\ emission
is seen in the first VA spectrum, but absorption is seen in the spectra
acquired 16 minutes later and on the next night.  No velocity corrections have
been applied to the VA spectra in order to preserve the relative velocities,
but the HD 38392 spectrum was shifted to roughly align it with HD 222259B.} 

\figcaption{The top spectrum is the first night's lightly smoothed first 
spectrum of the Li region of HD 222259A.  The bottom spectrum, 
offset vertically by an additive constant, is the spectrum of HD 222259B.
The Li line is of comparable or larger strength than the $\lambda$6717
\ion{Ca}{1} feature.}

\figcaption{The normalized index of Ca H and K emission, \lR, versus \bv\
color for the 624 stars observed in Paper I.  Here we have updated our
knowledge of multiplicity among the Very Active stars.  Multiple systems are
indicated by crosses, and stars that are probably single are circled.}

\figcaption{Color-magnitude diagram for stars with Hipparcos parallaxes.
Stars are identified by the letters in Table 2.
The squares represent stars that are probable or possible binaries,
while the circles represent stars that are probably single.  The dashed lines
are isochrones from \cite{siess97} (1997) corresponding to ages, from top to
bottom, of 1, 5, 10, 15, and 20 Myr.  The solid line at the bottom is
the ZAMS.  Some points have been offset slightly to avoid overlap.}

\end{document}